\documentclass[a4,12pt]{article}

\usepackage{graphicx}

\begin{document}

\begin{center}
{\large\bf Captures of Hot and Warm Sterile Antineutrino Dark Matter
\\ on EC-decaying $^{\bf 163}{\bf Ho}$ Nuclei}
\end{center}

\vspace{0.3cm}

\begin{center}
{\bf Y.F. Li} $^{\rm a}$ \footnote{E-mail: liyufeng@ihep.ac.cn} ~
and ~ {\bf Zhi-zhong Xing} $^{\rm a \ b}$
\footnote{E-mail: xingzz@ihep.ac.cn} \\
{$^{\rm a}$ Institute of High Energy Physics, Chinese Academy of
Sciences, Beijing 100049, China \\
$^{\rm b}$ Center for High Energy Physics, Peking University,
Beijing 100080, China}
\end{center}

\setcounter{footnote}{0}

\vspace{2.5cm}

\begin{abstract}
Capturing low-energy electron antineutrinos on radioactive
$^{163}{\rm Ho}$ nuclei, which decay into $^{163}{\rm Dy}$ via
electron capture (EC), is a noteworthy opportunity to detect relic
sterile antineutrinos. Such hypothetical particles are more or less
implied by current experimental and cosmological data, and they
might be a part of hot dark matter or a candidate for warm dark
matter in the Universe. Using the isotope $^{163}{\rm Ho}$ as a
target and assuming reasonable active-sterile antineutrino mixing
angles, we calculate the capture rate of relic electron
antineutrinos against the corresponding EC-decay background in the
presence of sterile antineutrinos at the sub-eV or keV mass scale.
We show that the signature of hot or warm sterile antineutrino dark
matter should in principle be observable, provided
the target is big enough and the energy resolution is good enough. \\

\hspace{-0.7cm}
\end{abstract}

\newpage

\section{Introduction}

Although the existence of dark matter (DM) in the Universe has been
established, what it is made of remains a fundamental puzzle
\cite{XZ}. Within the standard model (SM) three kinds of active
neutrinos and their antiparticles, whose masses lie in the sub-eV
range, may constitute hot DM. Beyond the SM one or more species of
sterile neutrinos and antineutrinos at a similar mass scale may also
form hot DM, if they were thermalized in the early Universe as their
active counterparts. Such light sterile particles are hypothetical,
but their existence is more or less implied by current experimental
and cosmological data. On the one hand, the long-standing LSND
antineutrino anomaly \cite{LSND}, the more recent MiniBooNE
antineutrino anomaly \cite{M} and the latest reactor antineutrino
anomaly \cite{R} can all be interpreted as the active-sterile
antineutrino oscillations in the assumption of two kinds of sterile
antineutrinos whose masses are close to 1 eV \cite{Schwetz}. On the
other hand, an analysis of the existing data on the cosmic microwave
background (CMB), galaxy clustering and supernovae Ia favors some
extra radiation content in the Universe and one or two species of
sterile neutrinos and antineutrinos at the sub-eV mass scale
\cite{Raffelt}
\footnote{If the Big Bang nucleosynthesis (BBN) bound is taken into
account, however, only one species of light sterile neutrinos and
antineutrinos is allowed \cite{Mangano}.}.
We are therefore open-minded to conjecture that hot DM may in
general consist of both active and sterile components. These relics
of the Big Bang form the unseen cosmic neutrino background (C$\nu$B)
and cosmic antineutrino background (C$\overline{\nu}$B), whose
temperature $T^{}_\nu$ is slightly lower than the CMB temperature
$T^{}_\gamma$ (i.e., $T^{}_\nu = \sqrt[3]{4/11} \ T^{}_\gamma \simeq
1.945$ eV today).

But hot DM only has a tiny contribution to the total matter density
of the Universe. A careful analysis of the structure formation
indicates that most DM should be cold (nonrelativistic) or warm
(semirelativistic) at the onset of the galaxy formation, when the
temperature of the Universe was about 1 keV \cite{PDG}. A number of
candidates for cold DM, such as weakly interacting massive particles
and axions \cite{Feng}, have so far been investigated. In
comparison, warm DM is another interesting possibility of accounting
for the observed non-luminous and non-baryonic matter content in the
Universe. Its existence may allow us to solve or soften several
problems that we have recently encountered in the DM simulations
\cite{Bode} (e.g., to damp the inhomogeneities on small scales by
reducing the number of dwarf galaxies or to smooth the cusps in the
DM halos). A good candidate for warm DM should be sterile neutrinos
and antineutrinos, if their masses are in the keV range and their
lifetimes are much longer than the age of the Universe
\cite{Review}. They could be produced in the early Universe in
several ways \cite{Shi}, and should be able to suppress the
formation of dwarf galaxies and other small-scale structures and
have impacts on the X-ray spectrum, the velocity distribution of
pulsars and the formation of the first stars \cite{Kusenko1}. Hence
it is likely to constrain their masses and mixing angles by
measuring the Lyman-$\alpha$ forest and X-ray fluxes
\cite{Abazajian}. In this connection some preliminary hints of keV
sterile neutrinos and antineutrinos have recently been discussed
\cite{WDM}.

On the theoretical side, there are some interesting models which can
accommodate sterile neutrinos and antineutrinos at either keV
\cite{keV} or sub-eV \cite{eV} mass scales. A model-independent
argument \cite{XING} is also supporting the conjecture of warm DM
hiding out in the ``flavor desert" of the SM fermion mass spectrum
\cite{LX11}. Here our main concern is purely phenomenological: how
can one directly probe the sterile component of the
C$\overline{\nu}$B and the keV sterile {\it antineutrino} DM? In
Ref. \cite{LLX} it has been shown that the sterile component of the
C$\nu$B can in principle be detected by means of the thresholdless
reaction $\nu^{}_e + ~^3{\rm H} \to ~^3{\rm He} + e^-$, because the
mass eigenstates of sub-eV sterile neutrinos contribute to
$\nu^{}_e$ and thus leaves a distinct imprint on the electron energy
spectrum when they are captured on $^3{\rm H}$ nuclei
\footnote{This approach was first proposed by Weinberg
\cite{Weinberg} and by Irvine and Humphreys \cite{Irvine} to detect
the active component of the C$\nu$B. It has recently attracted
more interest because it seems to be the most promising
possibility for the direct laboratory detection of relic neutrinos
(see Refs. \cite{LLX} and \cite{Cocco}---\cite{Hodak}).}.
The same idea has also been used to capture the keV sterile neutrino
DM on radioactive $\beta$-decaying $^3{\rm H}$ and $^{106}{\rm Ru}$
nuclei \cite{LX11,Liao}. However, this approach does not directly
apply to the capture of light sterile {\it antineutrinos}, simply
because it is $\nu^{}_e$ (instead of $\overline{\nu}^{}_e$) that is
involved in the capture reaction.

A possible way out is to make use of some radioactive nuclei which
can decay via electron capture (EC) \cite{Cocco2}. It has recently
been pointed out that the isotope $^{163}{\rm Ho}$, which undergoes
an EC decay into $^{163}{\rm Dy}$ with a small energy release ($Q
\simeq 2.5$ keV), may serve as a target to directly detect the
active component of the C$\overline{\nu}$B \cite{Lusignoli}. This
method also works in the presence of flavor effects, which are found
to be appreciable and even important in some cases \cite{LX11EC}.
Here we shall adopt the same idea to probe the {\it sterile}
component of the C$\overline{\nu}$B and the keV sterile {\it
antineutrino} DM. The EC decay of $^{163}{\rm Ho}$ \cite{Bambynek}
can be written as
\begin{equation}
^{163}{\rm Ho}  + e^-_{i (\rm shell)} \to  \; ^{163}{\rm Dy}^{*}_{i}
+ \nu^{}_e  \to  \; ^{163}{\rm Dy} + E^{}_i + \nu^{}_e \; ,
\end{equation}
and a thresholdless capture of the incoming $\overline{\nu}^{}_e$ on
the EC-decaying $^{163}{\rm Ho}$ may happen via
\begin{equation}
\overline{\nu}^{}_e + \, ^{163}{\rm Ho} + e^-_{i (\rm shell)} \to
\; ^{163}{\rm Dy}^{*}_{i} \to  \; ^{163}{\rm Dy} + E^{}_i \; ,
\end{equation}
where $e^-_{i \rm (shell)}$ is an orbital electron from the $i$-th
shell of $^{163}{\rm Ho}$, and $E^{}_i$ is the corresponding binding
energy of the electron hole in $^{163}{\rm Dy}$. Our main purpose is
to identify a signal shown in Eq. (2) from the background described
by Eq. (1) in the presence of relic sub-eV or keV sterile antineutrinos.
Such a study has not been done before and makes sense at least in
the following three aspects. First, it is a useful and nontrivial
extension of the discussions about the active component of the C$\overline{\nu}$B in Refs. \cite{Lusignoli} and \cite{LX11EC}
and may provide us with a novel approach towards the direct detection
of relic sterile antineutrinos, which should be compared with
relic sterile neutrinos. Second, it illustrates a common way to
probe hot and warm sterile antineutrino DM. Although this way is
extremely challenging and its prospect is rather remote,
it is hitherto the most promising way in this connection.
Third, it may serve to highlight the
importance of doing a high-statistics {\it calorimetric} experiment
to measure the $^{163}{\rm Ho}$ spectrum, probe the absolute
neutrino mass scale \cite{Mare} and even detect the
C$\overline{\nu}$B and warm sterile antineutrino DM.

The remaining parts of this paper are organized as follows. In
section 2 we summarize the main formulas which can be used to
calculate the energy spectrum of the EC decay in Eq. (1) and the
rate of the relic antineutrino capture in Eq. (2). Section 3 is
devoted to the capture of hot sterile antineutrino DM in the (3+2)
scheme of active-sterile antineutrino mixing. We show that the
signature is located on the right-hand side of the spectral endpoint
of the EC decay, and their interval is detectable if the target is
big enough, the energy resolution is good enough and the
gravitational clustering of sub-eV sterile antineutrinos is
significant around the Earth. Section 4 is devoted to the capture of
warm sterile antineutrino DM on $^{163}{\rm Ho}$ nuclei. We
calculate the capture rate by assuming the existence of one species
of sterile antineutrinos at the keV mass scale. We find that it is
in principle possible to identify a signal of this kind of warm DM,
but the required target mass is too big to be accomplishable in the
foreseeable future. We conclude in section 5.

\section{Relic antineutrino captures}

The key point of relic antineutrino captures on the EC-dcaying
$^{163}{\rm Ho}$ nuclei is to capture the relic {\it electron}
antineutrinos no matter how low their kinetic energies are. In the
presence of $N^{}_{\rm s}$ species of light sterile neutrinos and
antineutrinos, the flavor eigenstates $|\nu^{}_e\rangle$ and
$|\overline{\nu}^{}_e\rangle$ can be written as
\begin{eqnarray}
\left|\nu^{}_e \right\rangle \hspace{-0.2cm} & = & \hspace{-0.2cm}
\sum_k V^*_{e k} \left|\nu^{}_k \right\rangle \; ,
\nonumber \\
\left|\overline{\nu}^{}_e \right\rangle \hspace{-0.2cm} & = &
\hspace{-0.2cm} \sum_k V^{}_{e k} \left| {\overline{\nu}^{}_k}
\right\rangle \; ,
\end{eqnarray}
where $|\nu^{}_k\rangle$ (or $|\overline{\nu}^{}_k\rangle$) stands
for the mass eigenstate of an active (for $1\leq k \leq 3$) or
sterile (for $4 \leq k \leq N^{}_{\rm s}$) neutrino (or
antineutrino), and $V^{}_{e k}$ denotes an element in the first row
of the $(3+N^{}_{\rm s})\times (3+N^{}_{\rm s})$ neutrino mixing
matrix $V$ \cite{GR}. Given current experimental constraints on
sterile antineutrinos \cite{Schwetz}, it is reasonable to assume
that the light sterile antineutrinos under consideration do not
significantly affect the values of two mass-squared differences and
three mixing angles of three active antineutrinos extracted from
solar, atmospheric, reactor and accelerator neutrino oscillation
data \cite{PDG}. In this assumption we shall use $\Delta m^2_{21}
\simeq 7.6 \times 10^{-5} ~{\rm eV}^2$ and $|\Delta m^2_{31}| \simeq
2.4 \times 10^{-3} ~{\rm eV}^2$ together with $\theta^{}_{12} \simeq
34^\circ$ and $\theta^{}_{13} \simeq 10^\circ$ as typical inputs for
our analysis. Depending on the sign of $\Delta m^2_{31}$, there are
two possible mass patterns for active antineutrinos: $m^{}_1 <
m^{}_2 < m^{}_3$ (normal hierarchy) or $m^{}_3 < m^{}_1 < m^{}_2$
(inverted hierarchy). In either case the absolute mass scale is
unknown, but its upper bound is expected to be of ${\cal O}(0.1)$ eV
as constrained by current cosmological data \cite{WMAP10}. We shall
specify the values of $m^{}_k$ and $|V^{}_{ek}|$ (for $k = 4,
\cdots, N^{}_{\rm s}$) when we numerically calculate the capture
rates of relic sterile antineutrinos in sections 3 and 4.

Now let us consider the EC decay of $^{163}{\rm Ho}$ in Eq. (1). If
the $Q$-value of this reaction is defined as the mass difference
between $^{163}{\rm Ho}$ and $^{163}{\rm Dy}$, then the energy
spectrum of the outgoing neutrinos will be given by a series of
lines at $Q-E^{}_i$. For the time being $Q$ has been constrained to
the range $2.3 ~{\rm keV} \leq Q \leq 2.8 ~{\rm keV}$
\cite{Lusignoli}, and we shall typically take $Q \simeq 2.5$ keV in
our numerical calculations. Taking account of the Breit-Wigner
resonance form of the atomic levels, one has obtained the energy
spectrum of the detected EC events \cite{Bambynek} and its integral
in the narrow-width approximation \cite{Rujula}. As for the
thresholdless capture of the incoming relic antineutrinos on the
EC-decaying $^{163}{\rm Ho}$ nuclei in Eq. (2), the de-excitation
energy of unstable $^{163}{\rm Dy}^{*}_{i}$ is in principle
monoenergetic for each antineutrino mass eigenstate
$\overline{\nu}^{}_k$ (i.e., $T^{}_k \equiv
E^{}_{\overline{\nu}^{}_k} + Q$). Convoluted with a finite energy
resolution in a realistic experiment, the ideally discrete energy
lines of the final states in Eq. (2) must spread and then form a
continuous spectrum. As usual, we adopt a Gaussian energy resolution
function defined by
\begin{equation}
R(T, \, T^{}_k) = \frac{1}{\sqrt{2\pi} \,\sigma} \exp\left[-\frac{(T
- T^{}_k)^2}{2\sigma^2} \right] \; ,
\end{equation}
where $T^{}$ is the overall energy of an event detected in the
experiment. Using $\Delta$ to denote the experimental energy
resolution (i.e., the full width at half maximum of a Gaussian
energy resolution for the detected events), we have $\Delta =
2\sqrt{2\ln 2} \,\sigma \simeq 2.35482 \,\sigma$. Then the
differential antineutrino capture rate reads \cite{LX11EC}
\begin{eqnarray}
\frac{{\rm d} \lambda^{}_{\overline{\nu}}}{{\rm d} T} = {G_{\beta}^2
\over 2} \sum_i \sum_k n^{}_{\overline{\nu}^{}_k} |V^{}_{ek}|^2 R(T,
\, T_k) \, n^{}_i \, C^{}_i  \, \beta_i^2 \, B^{}_i \, {\Gamma^{}_i
\over 2\pi} \cdot {1 \over (T^{}_k-E^{}_i)^2 +\Gamma_i^2/4} \; ,
\end{eqnarray}
where $G^{}_{\beta} \equiv G^{}_{\rm F} \cos \theta^{}_{\rm C}$ with
$\theta^{}_{\rm C} \simeq 13^\circ$ being the Cabibbo angle of quark
flavor mixing, $n^{}_{\overline{\nu}^{}_k}$ denotes the number
density of $\overline{\nu}^{}_k$, $n^{}_i$ is the fraction of
occupancy of the $i$-th electron shell, $C^{}_i$ stands for the
nuclear shape factor, $\beta^{}_i$ represents the Coulomb amplitude
of the electron radial wave function, $B^{}_i$ is an atomic
correction for the electron exchange and overlap, and $\Gamma^{}_i$
denotes the finite natural width of the $i$-th atomic level. The
$Q$-value of the EC decay in Eq. (1) is so small that only those
electrons from $M^{}_1$, $M^{}_2$, $N^{}_1$, $N^{}_2$, $O^{}_1$,
$O^{}_2$ and $P^{}_1$ levels can be captured \cite{Lusignoli}. In
accordance with Eq. (5), the energy spectrum of the EC decay should
also be convoluted with the Gaussian energy resolution
\cite{LX11EC}:
\begin{eqnarray}
{{\rm d} \lambda^{}_{\rm EC}\over {\rm d} T^{}} \hspace{-0.2cm} & =
& \hspace{-0.2cm} \int_0^{Q - {\rm min}(m^{}_k)} {\rm d} T^{}_{\rm
c} \, \left[ {G^{2}_{\beta} \over {4 \pi^2}} \, R(T, \, T^{}_{\rm
c}) \, (Q-T^{}_{\rm c}) \right .
\nonumber \\
&& \hspace{-0.2cm} \left . \times \sum_k
|V^{}_{ek}|^2\sqrt{(Q-T^{}_{\rm c})^2-m_k^2}~ \Theta(Q - T^{}_{\rm
c} - m^{}_k) \right .
\nonumber \\
&& \hspace{-0.2cm} \left . \times \sum_i n^{}_i \, C^{}_i \,
\beta_i^2 \, B^{}_i \, {\Gamma^{}_i \over 2\pi} \cdot {1 \over
(T^{}_{\rm c}-E^{}_i)^2+\Gamma_i^2/4} \right] \; ,
\end{eqnarray}
where the theta function $\Theta(Q - T^{}_{\rm c} - m^{}_k)$ has
been introduced to ensure the kinematic requirement. One may use
Eqs. (5) and (6) to calculate the rate of the relic antineutrino
capture on the EC-decaying $^{163}{\rm Ho}$ nuclei against the
corresponding background (i.e., the EC decay itself) in the presence
of active and sterile flavor effects, which are characterized by
both the antineutrino masses $m^{}_k$ and the antineutrino mixing
matrix elements $|V^{}_{ek}|$ (for $k=1,2, \cdots N^{}_{\rm s}$). It
is therefore possible to probe the existence of relic sterile
antineutrinos at either sub-eV or keV mass scales.

In our numerical calculations we shall input the values of those
parameters relevant to the atomic levels of $^{163}{\rm Dy}$ given
in Ref. \cite{Lusignoli} and references therein. The energy levels
of the captured electrons are assumed to be fully occupied (i.e.,
$n^{}_i \simeq 1$) and their binding energies and widths can be
found in Table 1 of Ref. \cite{Lusignoli}. The atomic corrections
for the electron exchange and overlap are neglected (i.e.,
$B_{i}\simeq 1$), and the ratios of the squared wave functions at
the origin (i.e., $\beta^{2}_{i}/\beta^2_{\rm M^{}_1}$) can be found
in Table 2 of Ref. \cite{Lusignoli}. Because the nuclear shape
factors $C^{}_i$ are approximately identical in an allowed
transition \cite{Bambynek}, they can be factored out from the sum in
Eqs. (5) and (6). Note that the numerical results of
$\lambda^{}_{\overline{\nu}}$ and $\lambda^{}_{\rm EC}$ can be
properly normalized by using the half-life of $^{163}{\rm Ho}$ via
the relation $\lambda^{}_{\rm EC}T^{}_{1/2}=\ln 2$, where
$T^{}_{1/2} \simeq 4570$ yr. Then the distributions of the number of
signal events and the number of background events are expressed,
respectively, as
\begin{eqnarray}
\frac{{\rm d} N^{}_{\rm S}}{{\rm d} T} \hspace{-0.2cm} & = &
\hspace{-0.2cm} {1 \over \lambda^{}_{\rm EC}} \cdot {{\rm d}
\lambda^{}_{\overline{\nu}} \over {\rm d} T} \cdot {\ln 2 \over
T^{}_{1/2}} \, N^{}_{\rm T} \, t \; ,
\nonumber \\
\frac{{\rm d} N^{}_{\rm B}}{{\rm d} T} \hspace{-0.2cm} & = &
\hspace{-0.2cm} {1 \over \lambda^{}_{\rm EC}} \cdot {{\rm d}
\lambda^{}_{\rm EC}\over {\rm d} T} \cdot {\ln 2 \over T^{}_{1/2}}
\, N^{}_{\rm T} \, t \;
\end{eqnarray}
for a given target factor $N^{}_{\rm T}$ (i.e., the number of
$^{163}{\rm Ho}$ atoms of the target) and for a given exposure time
$t^{}_{}$ in the experiment.

\section{Hot sterile antineutrino DM}

To be explicit, let us focus on the capture of hot sterile
antineutrino DM in the (3+2) flavor mixing scheme with two species
of sub-eV sterile neutrinos and antineutrinos \cite{Schwetz}. The
absolute mass scale of three active antineutrinos is characterized
by $m^{}_1$ (normal hierarchy) or $m^{}_3$ (inverted hierarchy), and
its value is typically taken to be 0.0 eV, 0.05 eV or 0.1 eV in our
numerical analysis. In addition, we choose $m^{}_4 \simeq 0.2$ eV
and $m^{}_5 \simeq 0.4$ eV together with $\theta^{}_{14} \simeq
\theta^{}_{15} \approx 10^\circ$ for two sterile antineutrinos
\footnote{The values of two sterile antineutrino masses taken here
are more favored by current cosmological data \cite{Raffelt}, but
they are somewhat smaller than those values extracted from a global
fit of the LSND, MiniBooNE and reactor antineutrino anomalies
\cite{Schwetz}. It is simply a matter of taste for us to take the
cosmological hints on sub-eV sterile neutrinos and antineutrinos
more seriously. Since the chosen values of $m^{}_4$ and $m^{}_5$ are
mainly for the purpose of illustration, they do not qualitatively
affect our conclusions.}.
Accordingly, we have $|V^{}_{e1}| \simeq 0.792$, $|V^{}_{e2}| \simeq
0.534$, $|V^{}_{e3}| \simeq 0.168$, $|V^{}_{e4}| \simeq 0.171$ and
$|V^{}_{e5}| \simeq 0.174$. Such sterile antineutrinos are expected
to be cosmologically friendly \cite{Raffelt}, and they should stay
in full thermal equilibrium in the early Universe \cite{Hannestad}.
In this case the standard Big Bang model predicts the average number
density $\langle n^{}_{\nu^{}_k} \rangle \simeq \langle
n^{}_{\overline{\nu}^{}_k} \rangle \simeq 56 ~{\rm cm}^{-3}$ today
for each species of active and sterile neutrinos and antineutrinos.

For our purpose, however, we are mainly concerned about the number
density of relic antineutrinos around the Earth because its value
$n^{}_{\overline{\nu}^{}_k}$ may be more or less enhanced by the
gravitational clustering effect. A detailed analysis of
gravitational clustering of relic neutrinos and antineutrinos in our
local neighborhood has been done in Ref. \cite{Wong}. Here we take
account of  this effect by conjecturing a simplified power law
relation between the relic antineutrino overdensity parameter
$\zeta^{}_k \equiv n^{}_{\overline{\nu}^{}_k}/\langle
n^{}_{\overline{\nu}^{}_k} \rangle$ and the corresponding
antineutrino mass $m^{}_k$ (for $k=1,2,\cdots,N^{}_{\rm s}$):
\begin{equation}
\zeta^{}_k \simeq 1 + A^{}_{} \left(\frac{m^{}_k}{1 ~{\rm eV}}
\right)^\Omega \; ,
\end{equation}
where $A$ and $\Omega$ are independent of the subscript $k$. Note
that the size of $\zeta^{}_k$ for a given value of $m^{}_k$ should
lie between the results obtained in two extreme cases as summarized
in Table 2 of Ref. \cite{Wong}. For simplicity, here we take the
average of every pair of those extreme values as the central input
of $\zeta^{}_k$ and one half of their difference as the error bar of
$\zeta^{}_k$. Then we have $\zeta^{}_k = \{1.5 \pm 0.1, 3.75 \pm
0.65, 8.2 \pm 1.8, 16 \pm 4\}$ corresponding to $m^{}_k = \{0.15,
0.3, 0.45, 0.6\}$ eV extracted from Table 2 in Ref. \cite{Wong}.
After a least square analysis of the relation between $m^{}_k$ and
$\zeta^{}_k$ as illustrated in Fig. 1, we find that the best-fit
values of $A$ and $\Omega$ are $\log_{10} A \simeq 1.7$ and $\Omega
\simeq 2.4$. One may easily see that the deviation of $\zeta^{}_k$
from one becomes significant only when $m^{}_k$ is larger than 0.1
eV (e.g., $\zeta^{}_k \simeq 2$ for $m^{}_k \simeq 0.2$ eV). So in
the (3+2) flavor mixing scheme under discussion, only two sub-eV
sterile antineutrinos are relatively sensitive to the gravitational
clustering effect.

We calculate the rate of the relic antineutrino capture on
$^{163}{\rm Ho}$ nuclei against the rate of the corresponding EC
decay by means of Eqs. (5), (6) and (7). Our numerical results are
presented in Figs. 2 and 3. Some discussions are in order.

(1) Fig. 2 illustrates the relic antineutrino capture rate as a
function of the overall energy release $T$ in the case of $\Delta
m^2_{31} >0$. In the left panel $\zeta^{}_k = 1$ (i.e.,
$n^{}_{\overline{\nu}^{}_k} = \langle n^{}_{\overline{\nu}^{}_k}
\rangle$) is assumed, and in the right panel the gravitational
clustering effect (i.e., $\zeta^{}_k >1$) has roughly been taken
into account with the help of Eq. (8). The impact of different
values of $m^{}_1$ on the capture rate can be vertically seen in
each panel. Note that the value of the finite energy resolution
$\Delta$ is taken in such a way that only the signal of hot {\it
sterile} antineutrino DM can be observed, since the signal of hot
{\it active} antineutrino DM has already been discussed in Ref.
\cite{LX11EC}. We find that it is in principle possible to
distinguish the signal from the background when $\Delta$ is smaller
than $0.1$ eV. As the value of $m^{}_1$ increases from 0 to 0.1 eV,
the signal curve moves towards the higher $T^{}_{} - Q^{}_{}$ region
while the background curve moves towards the lower $T^{}_{} -
Q^{}_{}$ region. Hence the interval between the peak of a signal and
the background becomes larger for a larger value of $m^{}_1$, making
it easier to detect this signal. If the heights of two neighboring
signals are quite different and their locations satisfy
$m^{}_{i}-m^{}_{j} \sim \Delta$, it will be difficult to distinguish
between them because the higher signal is much broader and may
essentially overwhelm the lower one. This feature can be seen in
Fig. 2(e) and Fig. 2(f), where the signal of $\overline{\nu}^{}_{4}$
and the signal of three active antineutrinos with nearly degenerate
masses merge into a single one. In this case only the signal of much
heavier $\overline{\nu}^{}_5$ is clearly distinguishable. In the
right panel of Fig. 2 one may observe that the gravitational
clustering effect can enhance the capture rate, and this effect is
more significant for heavier sterile antineutrinos. If the
gravitational clustering of relic antineutrinos were much more
significant around the Earth, it would be very helpful for us to
detect the C$\overline{\nu}$B by means of the capture reaction under
consideration.

(2) Fig. 3 shows the numerical results obtained from an analogous
analysis of the relic antineutrino capture rate in the case of
$\Delta m^2_{31} <0$. We see that it is quite similar to Fig. 2, and
their main differences appear in the location of the signal of three
active antineutrinos and that of the endpoint of the EC-decay
background. The first difference arises from the fact that the
location of the signal of active antineutrinos is dominated by the
mass eigenstates $\overline{\nu}^{}_{1}$ and $\overline{\nu}^{}_{2}$
which have slightly larger eigenvalues in the $\Delta m^2_{31} <0$
case than in the $\Delta m^2_{31} >0$ case. As for the second
difference, the spectral endpoint of the EC decay of $^{163}{\rm
Ho}$ is located at a smaller $T^{}_{} - Q^{}_{}$ region in the
$\Delta m^2_{31} <0$ case, simply because the lightest mass
eigenstate $\overline{\nu}^{}_{3}$ hiding out in
$\overline{\nu}^{}_{e}$ is associated with the smallest active
antineutrino mixing matrix element $|V^{}_{e3}|$. So we conclude
that it is somewhat easier to observe the signals of hot active and
sterile antineutrino DM when three active antineutrinos have an
inverted mass hierarchy.

(3) Besides the energy resolution, another factor that obviously
affects the observability of hot antineutrino DM is the absolute
capture rate which depends on the number density of relic
antineutrinos $n^{}_{\overline{\nu}^{}_k}$, the flavor mixing matrix
elements $|V^{}_{ek}|$ and the number of the target particles
$N^{}_{\rm T}$. By using the default values of $|V^{}_{ek}|$ taken
above, we illustrate the iso-rate curves for the sterile
antineutrino overdensity parameter $\zeta^{}_5$ versus the target
mass in Fig. 4
\footnote{Instead of using Eq. (8), here we treat $\zeta^{}_5$ as a
free parameter independent of the neutrino masses $m^{}_k$. In the
literature there are some different neutrino overdensity models
which can actually predict very large values of $\zeta^{}_k$ (see,
e.g., Ref. \cite{Wigmans}).}.
Because of $|V^{}_{e4}| \sim |V^{}_{e5}|$ and $m^{}_4 \sim m^{}_5
\ll Q-E^{}_i$, one may obtain very similar iso-rate curves for
$\zeta^{}_4$ versus the target mass. These curves are simply
straight lines in the double logarithmic scale, which can easily be
understood with the help of Eqs. (5) and (7). Given $\zeta^{}_k =
10$, for example, a target with $10^2$ kg (or $10^3$ kg) $^{163}{\rm
Ho}$ could be enough to give a capture rate of one event per year
(or ten events per year).

Finally, let us give a brief comment on the properties of the
capture signals of hot sterile antineutrinos against the
corresponding EC decays by taking the masses of sterile
antineutrinos to be consistent with the values of $\Delta m^2_{41}$
and $\Delta m^2_{51}$ extracted from a global fit of the LSND,
MiniBooNE and reactor antineutrino anomalies (i.e., $\Delta
m^{2}_{41}\simeq 0.47\;{\rm eV}^2$ and $\Delta m^{2}_{51}\simeq
0.87\;{\rm eV}^2$ \cite{Schwetz}). In this case the dependence of
the capture rates on the energy resolution, antineutrino number
densities and target factors is quite similar to the one discussed
above. The only change is the location of each sterile antineutrino
signal. Because of $m^{}_4 \simeq \sqrt{\Delta m^2_{41}} \simeq
0.69$ eV and $m^{}_5 \simeq \sqrt{\Delta m^2_{51}} \simeq 0.93$ eV
in the present situation, it will be much easier to identify the
sterile antineutrino signals on EC-decaying $^{163}{\rm Ho}$ nuclei
if the magnitudes of the mixing matrix elements $|V^{}_{e4}|$ and
$|V^{}_{e5}|$ keep unchanged.

\section{Warm sterile antineutrino DM}

Now we turn to the capture of warm sterile antineutrino DM in the
(3+1) flavor mixing scheme with one species of keV sterile neutrinos
and antineutrinos. To be explicit, we fix $m^{}_4 \simeq 2.0$ keV
and $|V^{}_{e4}|^{2} \simeq 5.0\times 10^{-7}$ in our numerical
analysis, just for the purpose of illustration \cite{LX11}. Because
our main concern is to observe a possible signal of the relic
antineutrino capture in the keV mass region, we simply assume three
active antineutrinos to have a normal mass hierarchy with $m^{}_1 =
0$. A very similar signal can also be obtained for the inverted or
nearly degenerate mass pattern of three active antineutrinos. One
may also consider the gravitational clustering effect on the
C$\overline{\nu}$B as described in Eq. (8), but it is insignificant
in the present scenario where only active antineutrinos contribute
to the C$\overline{\nu}$B. As for the keV sterile neutrinos and
antineutrinos, we follow Ref. \cite{Shi} to assume that they were
produced in the early Universe through active-sterile neutrino or
antineutrino oscillations and their number densities are able to
account for the total amount of DM. Given the average density of DM
in our Galactic neighborhood (i.e., $\rho^{\rm local}_{\rm DM}
\simeq 0.3 ~{\rm GeV} \ {\rm cm}^{-3}$ \cite{Kamionkowski}), it is
straightforward to estimate the number density of $\nu^{}_4$ or
$\overline{\nu}^{}_4$. We obtain $n^{}_{\nu^{}_4} \simeq
n^{}_{\overline{\nu}^{}_4} \simeq 5 \times 10^{4} \ (3 ~{\rm
keV}/m^{}_4) ~{\rm cm}^{-3}$.

We calculate the capture rate of warm sterile antineutrino DM on
$^{163}{\rm Ho}$ nuclei against the corresponding EC-decay
background by using Eqs. (5), (6) and (7). Our numerical result is
shown in Fig. 5. Some discussions are in order.

(1) In obtaining Fig. 5 we have chosen a typical value of the finite
energy resolution $\Delta$ to distinguish the signal from the
background. The endpoint of the background is sensitive to $\Delta$,
while the peak of the signal is always located at $T^{}_{} = Q^{}_{}
+ m^{}_4$. So a comparison between $\Delta$ and $m^{}_4$ can easily
reveal the signal-to-background ratio. The required energy
resolution to identify the signal of warm sterile antineutrino DM is
of ${\cal O}(0.1) ~{\rm keV}$, which should be easily reachable in a
realistic experiment.

(2) The main problem which makes the observability of keV sterile
antineutrino DM rather dim and remote is the tiny capture rate. The
latter is strongly suppressed for two simple reasons. On the one
hand, the mixing factor between three active antineutrinos and the
keV sterile antineutrino is too small \cite{Review}. On the other
hand, the Breit-Wigner distribution function is too small when the
signal of $\overline{\nu}^{}_4$ is located far away from the
resonance energies $E^{}_{i}$. In Fig. 6 we depict the iso-rate
curves for the mixing matrix element $|V^{}_{e4}|^2$ versus the
target mass. One can see that a target with 600 ton $^{163}{\rm Ho}$
is needed for $|V^{}_{e4}|^2 \sim {\cal O}(10^{-6})$, so as to
obtain a capture rate of one event per year. Hence it is almost
hopeless to capture warm sterile antineutrino DM on the EC-decaying
$^{163}{\rm Ho}$ nuclei in the foreseeable future. In comparison, it
seems somewhat more hopeful to detect warm DM in the form of keV
sterile {\it neutrinos} on radioactive beta-decaying $^3{\rm H}$ and
$^{106}{\rm Ru}$ nuclei in the long term \cite{LX11}.

Finally, let us give some brief comments on the detection prospects
for three kinds of relic antineutrinos (i.e., hot active
antineutrino DM, hot sterile antineutrino DM and warm sterile
antineutrino DM). Given a target made of the isotope $^{163}{\rm
Ho}$, the capture of each kind of DM in a realistic experiment
crucially relies on its energy resolution and target mass. Because
three active antineutrinos have relatively large mixing angles and
relatively small masses, the corresponding hot active antineutrino
DM has the largest capture rate but requires the most stringent
energy resolution ($\Delta \simeq 0.015$ eV or smaller, as discussed
in Ref. \cite{LX11EC}) for us to distinguish a signal from its
background. In contrast, the situation for capturing warm sterile
antineutrino DM is just the opposite. Here $\Delta \sim {\cal
O}(0.1)$ keV is good enough because of $m^{}_4 \sim {\cal O}(1)$
keV, and this energy resolution requirement can be satisfied even
today. But the capture rate of $\overline{\nu}^{}_4$ on the
EC-decaying $^{163}{\rm Ho}$ nuclei is so small that the desired
target mass has to be formidably large. As we have seen in the above
analysis, the capture of hot sterile antineutrino DM at the sub-eV
(or eV) mass scale requires a relatively mild energy resolution and
a reasonably large target mass as compared with the situations for
detecting hot active antineutrino DM and warm sterile antineutrino
DM in the same way. So it might be possible to probe the sterile
component of the C$\overline{\nu}$B by using the isotope $^{163}{\rm
Ho}$ as a target in the future.

\section{Concluding remarks}

To pin down what DM is made of has been one of the most important
and most challenging problems in particle physics and cosmology. In
this paper we have addressed ourselves to the direct laboratory
detection of possible contributions of light sterile antineutrinos
to DM. Such hypothetical particles might be a part of hot DM if
their masses are in the sub-eV (or eV) range, or a good candidate
for warm DM if their masses lie in the keV range. Choosing the
isotope $^{163}{\rm Ho}$ as a target and assuming reasonable
active-sterile antineutrino mixing angles, we have calculated the
capture rate of relic electron antineutrinos against the
corresponding EC-decay background in the presence of sterile
antineutrinos at either sub-eV or keV mass scales. Our analysis
shows that the signature of hot or warm sterile antineutrino DM
should in principle be observable, provided the target is big
enough, the energy resolution is good enough and the gravitational
clustering effect is significant enough. We admit that our numerical
results are quite preliminary and mainly serve for illustration, but
we stress that such a direct laboratory search for hot and warm
sterile antineutrino DM is fundamentally important and deserves
further attention and more detailed investigations.

Why has one chosen the EC-decaying $^{163}{\rm Ho}$ nuclei as a
proper target for probing the antineutrino content of DM? The reason
is simply that the captures of relic neutrinos and antineutrinos
require different types of radioactive nuclei: $\nu^{}_e$ is
involved in the capture reactions on the beta-decaying $^3{\rm H}$
and $^{106}{\rm Ru}$ nuclei, while $\overline{\nu}^{}_e$ is
associated with the capture reactions on the EC-decaying $^{163}{\rm
Ho}$ nuclei. So it makes sense to look for another rather stable
isotope which can undergo an EC decay with a much lower $Q$-value,
in order to do a more promising experiment for the direct detection
of hot and warm antineutrino DM. Although the present experimental
techniques are unable to lead us to a guaranteed measurement of
relic neutrinos and antineutrinos in the near future, we might have
a chance to make a success of this great exploration in the long
term.

\vspace{0.5cm}

This work was supported in part by the China Postdoctoral Science
Foundation under grant No. 20100480025 (Y.F.L.) and in part by the
National Natural Science Foundation of China under grant No.
10875131 (Z.Z.X.).

\newpage

\begin{figure}[p!]
\begin{center}
\begin{tabular}{cc}
\includegraphics*[bb=18 18 286 228, width=0.65\textwidth]{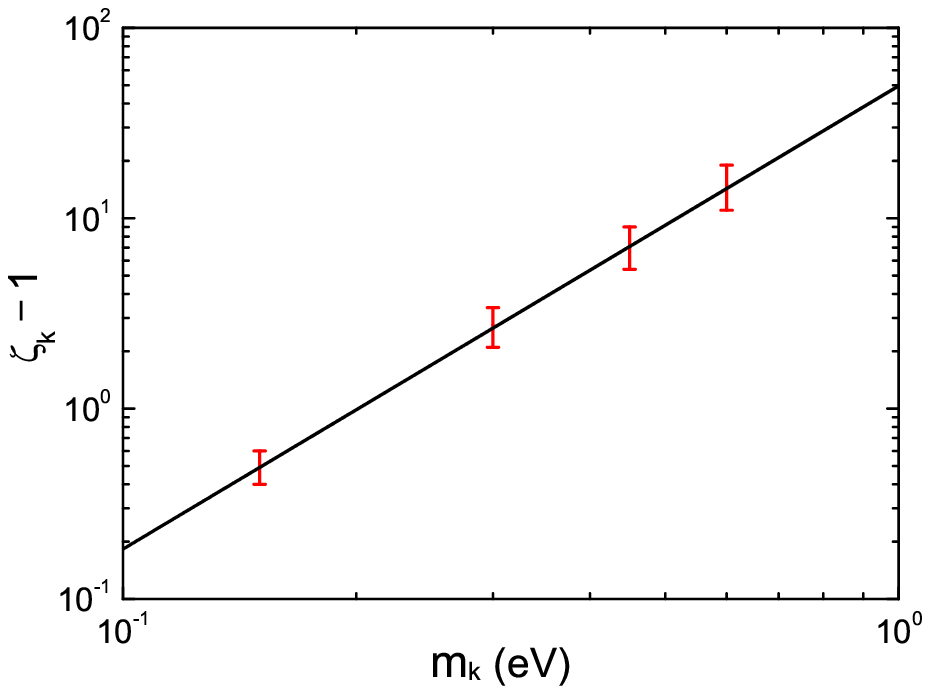}
\end{tabular}
\end{center}
\caption{The best-fit power law relation between the antineutrino
mass $m^{}_k$ and the overdensity parameter $\zeta^{}_k$
as a consequence of the gravitational clustering effect on relic
antineutrinos in our local neighborhood. The four points and their
error bars in this graph are extracted from Table 2 of Ref. \cite{Wong}.
}
\end{figure}

\begin{figure}[p!]
\begin{center}
\begin{tabular}{cc}
\includegraphics*[bb=18 18 276 210, width=0.46\textwidth]{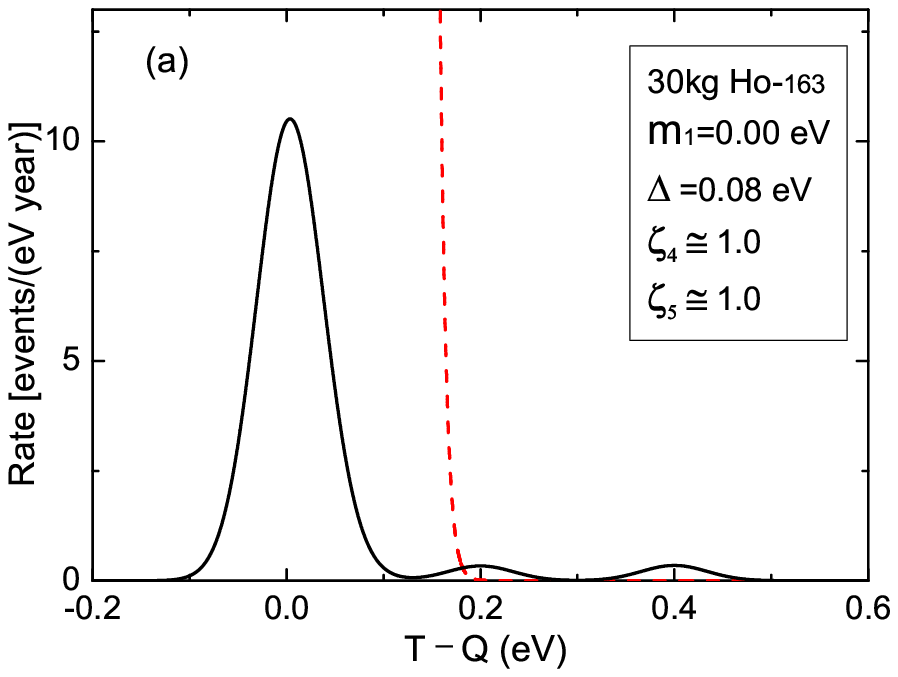}
&
\includegraphics*[bb=18 18 276 210, width=0.46\textwidth]{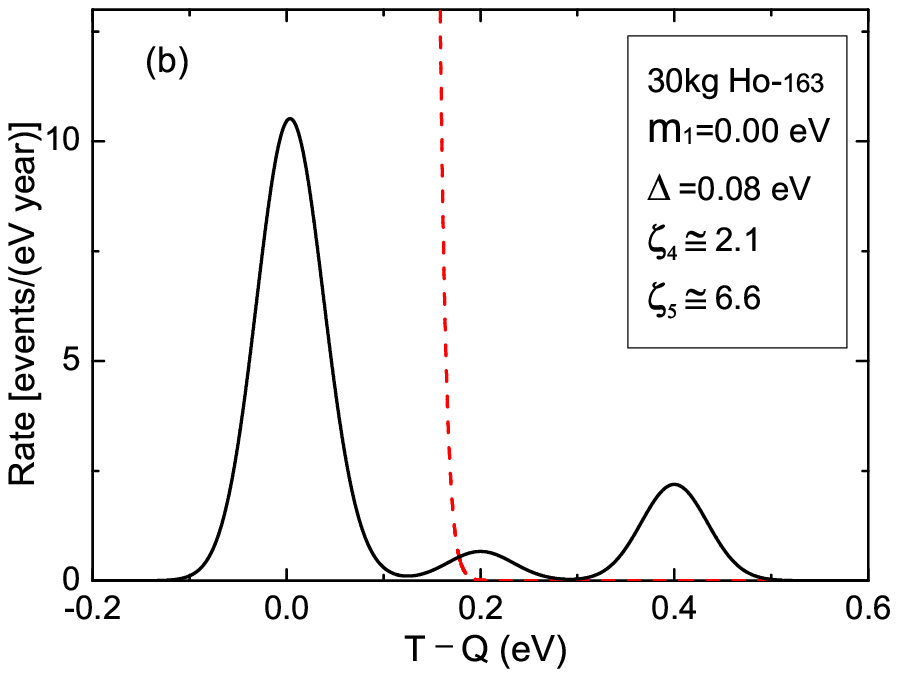}
\\
\includegraphics*[bb=18 18 276 220, width=0.46\textwidth]{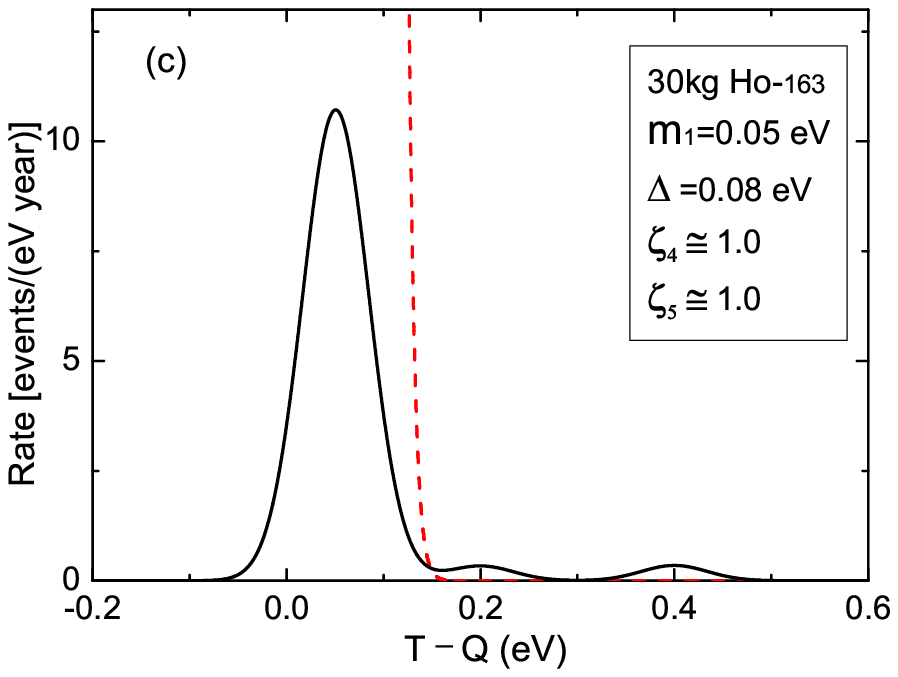}
&
\includegraphics*[bb=18 18 276 220, width=0.46\textwidth]{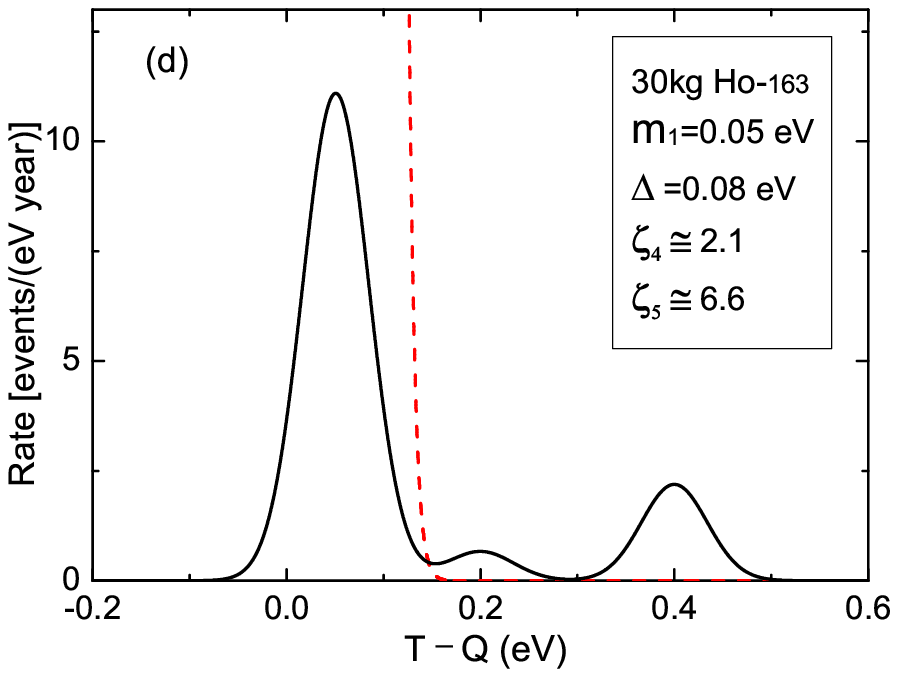}
\\
\includegraphics*[bb=18 18 276 220, width=0.46\textwidth]{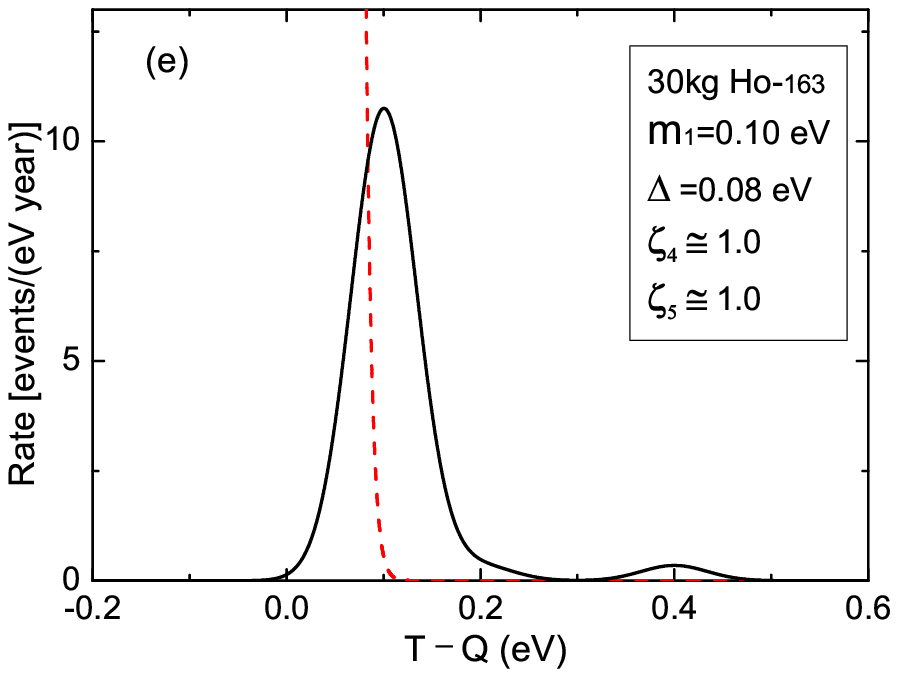}
&
\includegraphics*[bb=18 18 276 220, width=0.46\textwidth]{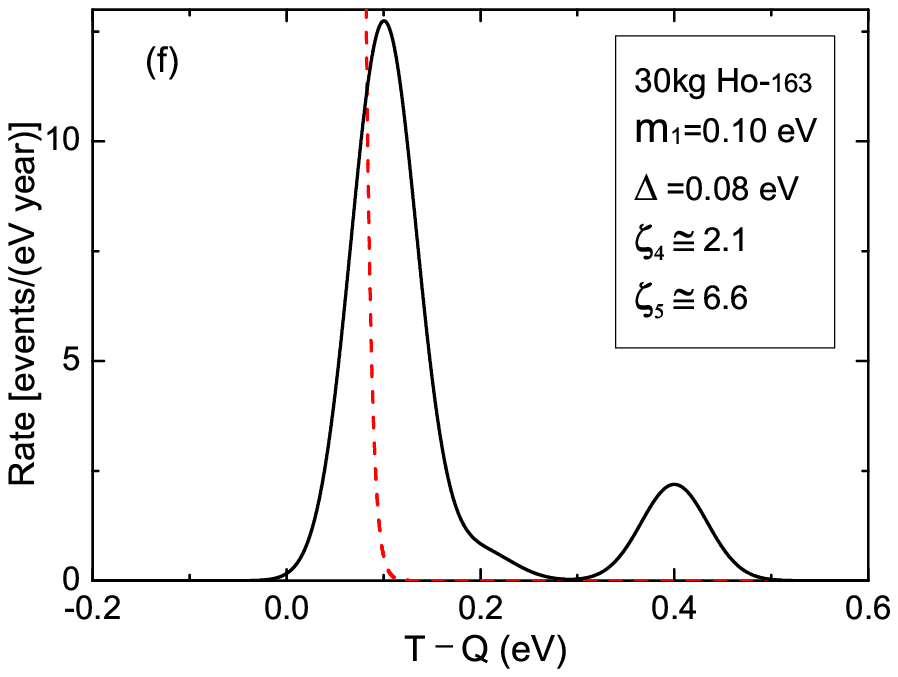}
\end{tabular}
\end{center}
\caption{The relic antineutrino capture rate as a function of the
overall energy release $T$ in the case of $\Delta m^2_{31} >0$,
where $\zeta^{}_k = 1$ in the left panel and $\zeta^{}_k > 1$
described by Eq. (8) in the right panel to illustrate possible
gravitational clustering effects on relic antineutrinos. The solid
and dashed curves represent the C$\overline{\nu}$B signature and its
background, respectively. The value of the finite energy resolution
$\Delta$ is taken in such a way that only the signal of the sterile
component of the C$\overline{\nu}$B can be seen.}
\end{figure}

\begin{figure}[p!]
\begin{center}
\begin{tabular}{cc}
\includegraphics*[bb=18 18 276 210, width=0.46\textwidth]{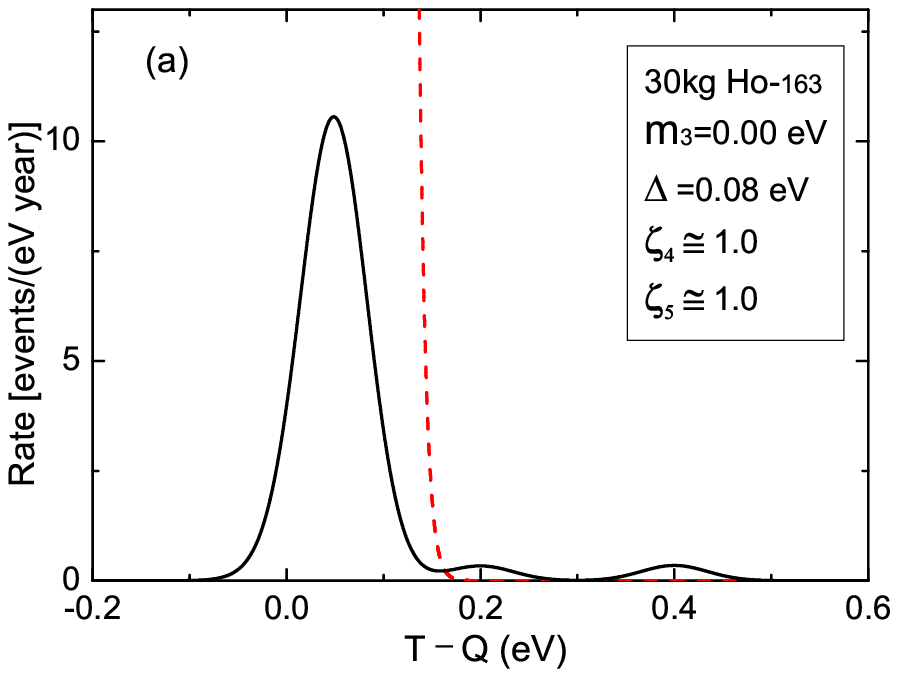}
&
\includegraphics*[bb=18 18 276 210, width=0.46\textwidth]{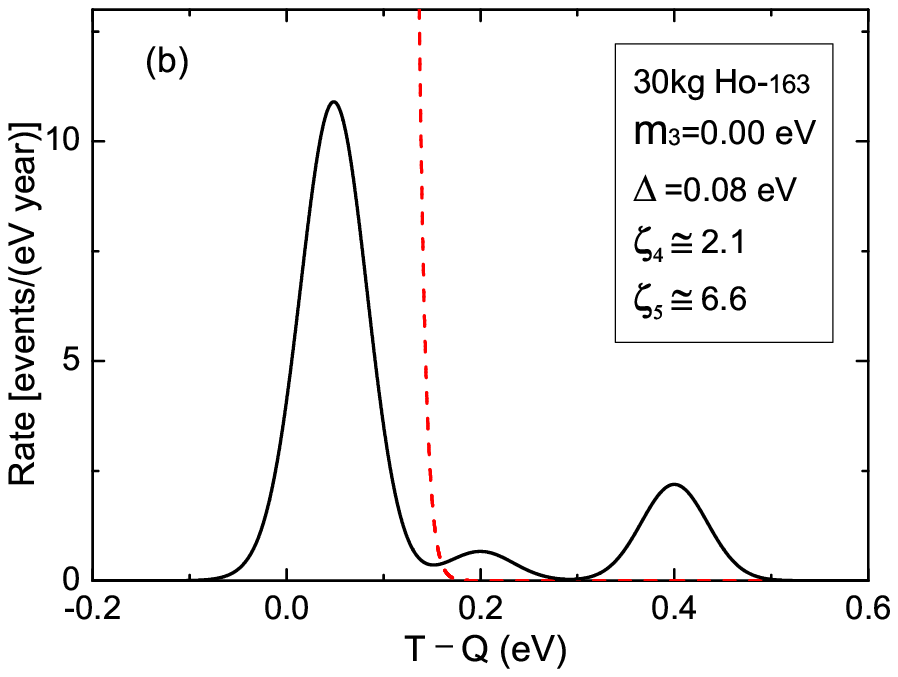}
\\
\includegraphics*[bb=18 18 276 220, width=0.46\textwidth]{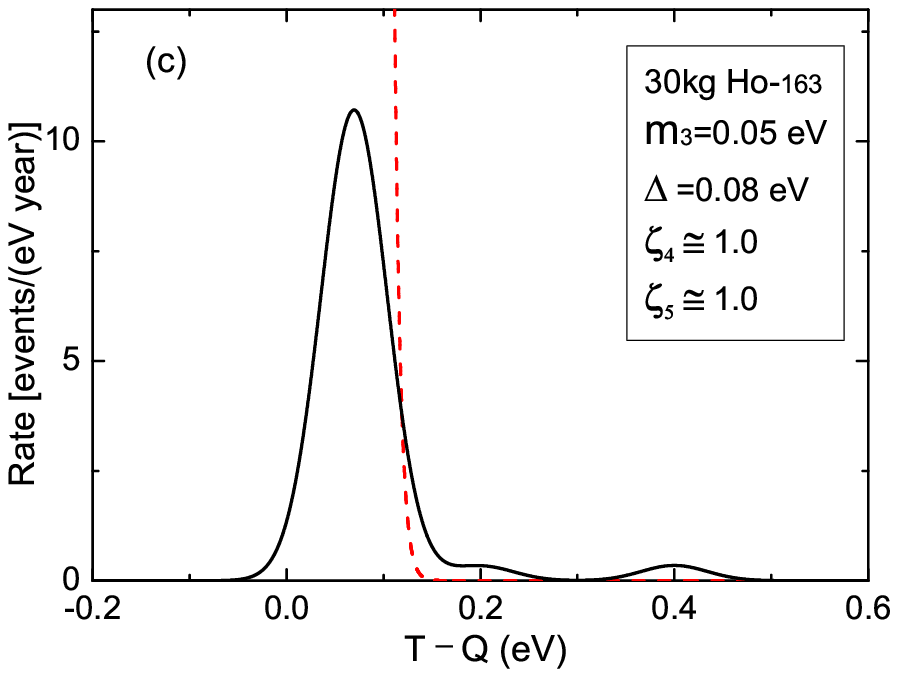}
&
\includegraphics*[bb=18 18 276 220, width=0.46\textwidth]{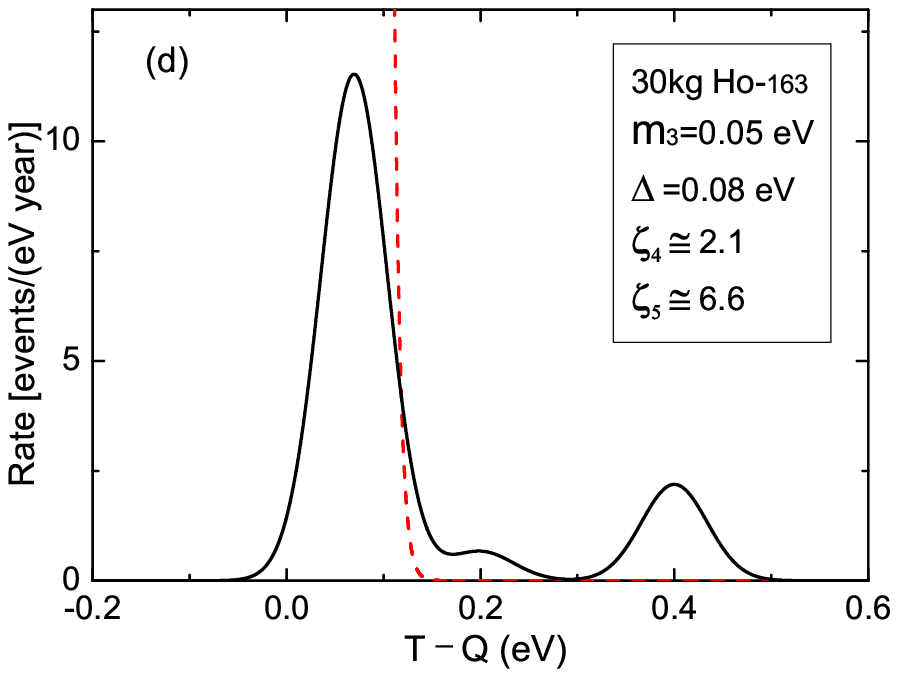}
\\
\includegraphics*[bb=18 18 276 220, width=0.46\textwidth]{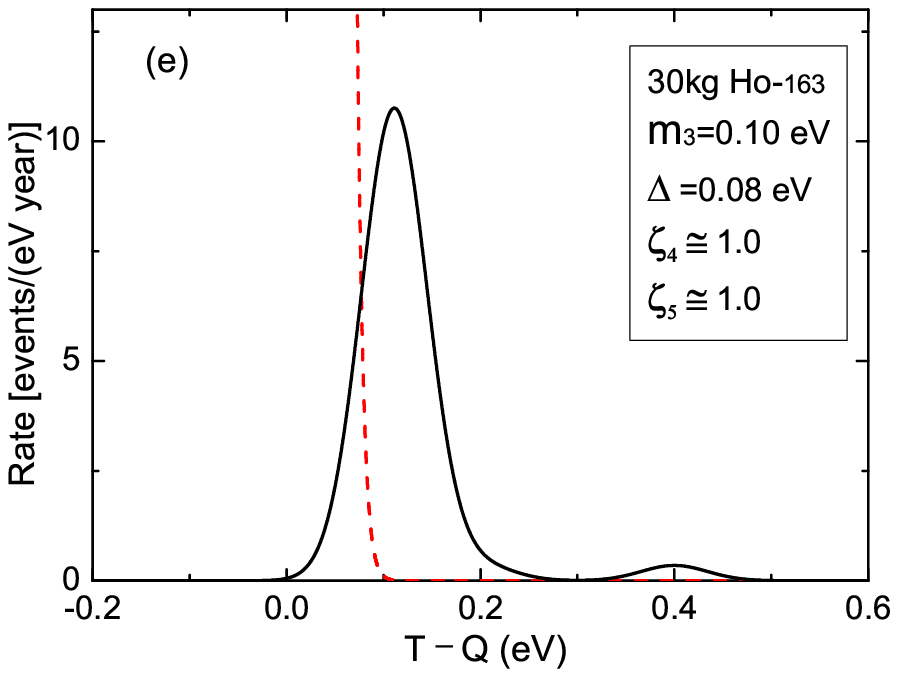}
&
\includegraphics*[bb=18 18 276 220, width=0.46\textwidth]{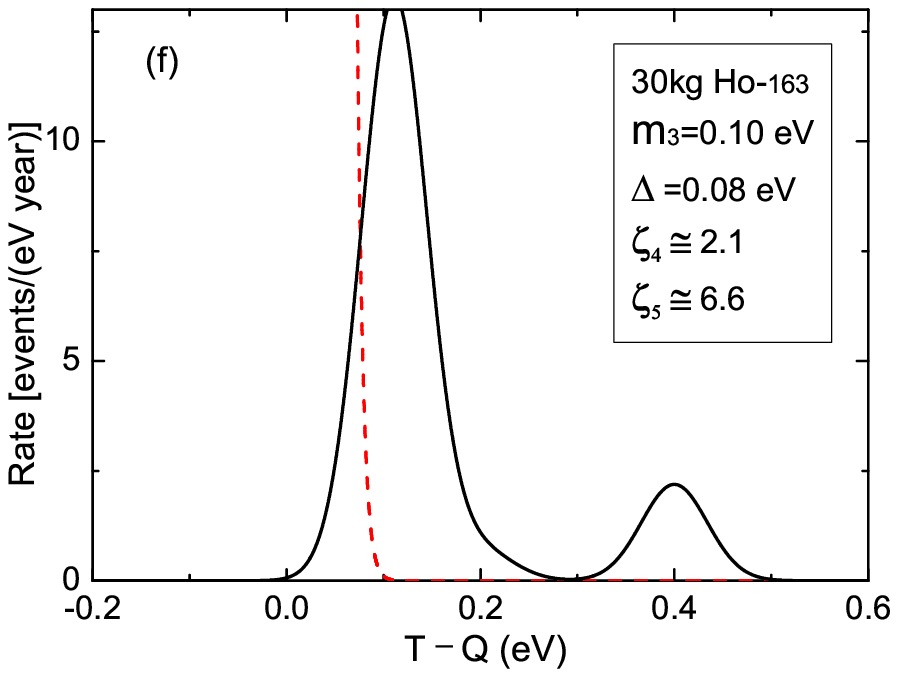}
\end{tabular}
\end{center}
\caption{The relic antineutrino capture rate as a function of the
overall energy release $T$ in the case of $\Delta m^2_{31} <0$,
where $\zeta^{}_k = 1$ in the left panel and $\zeta^{}_k > 1$
described by Eq. (8) in the right panel to illustrate possible
gravitational clustering effects on relic antineutrinos. The solid
and dashed curves represent the C$\overline{\nu}$B signature and its
background, respectively. The value of the finite energy resolution
$\Delta$ is taken in such a way that only the signal of the sterile
component of the C$\overline{\nu}$B can be seen.}
\end{figure}

\begin{figure}[p!]
\begin{center}
\begin{tabular}{c}
\includegraphics*[bb=18 16 284 230, width=0.65\textwidth]{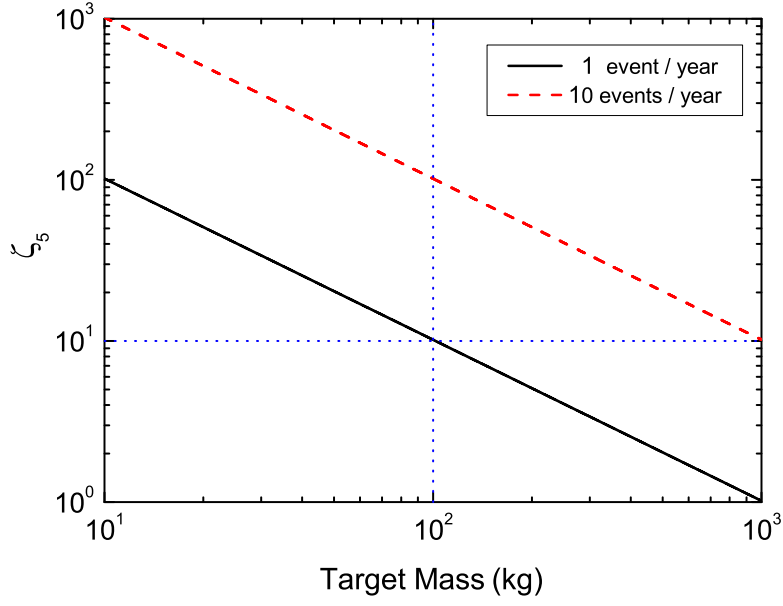}
\end{tabular}
\end{center}
\caption{The iso-rate curves of hot sterile antineutrino DM for the
overdensity parameter $\zeta^{}_5$ versus the target mass. Here
$\zeta^{}_5$ is treated as a free parameter and $m^{}_5 \simeq 0.4$
eV and $|V^{}_{e5}| \simeq 0.174$ have been input.}
\end{figure}

\begin{figure}[p!]
\begin{center}
\begin{tabular}{c}
\includegraphics*[bb=18 18 282 228, width=0.65\textwidth]{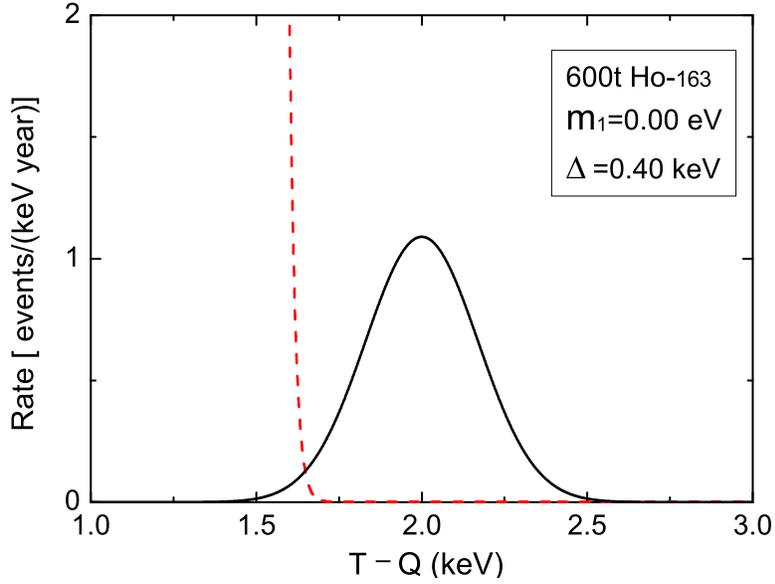}
\end{tabular}
\end{center}
\caption{The rate of capturing warm sterile antineutrino DM as a
function of the overall energy release $T$ in the (3+1) flavor
mixing scheme with $\Delta m^2_{31}>0$ and $m^{}_{4} \simeq 2\;{\rm
keV}$. The solid and dashed curves represent the signature and its
EC-decay background, respectively.}
\end{figure}

\begin{figure}[p!]
\begin{center}
\begin{tabular}{c}
\includegraphics*[bb=20 16 286 230, width=0.65\textwidth]{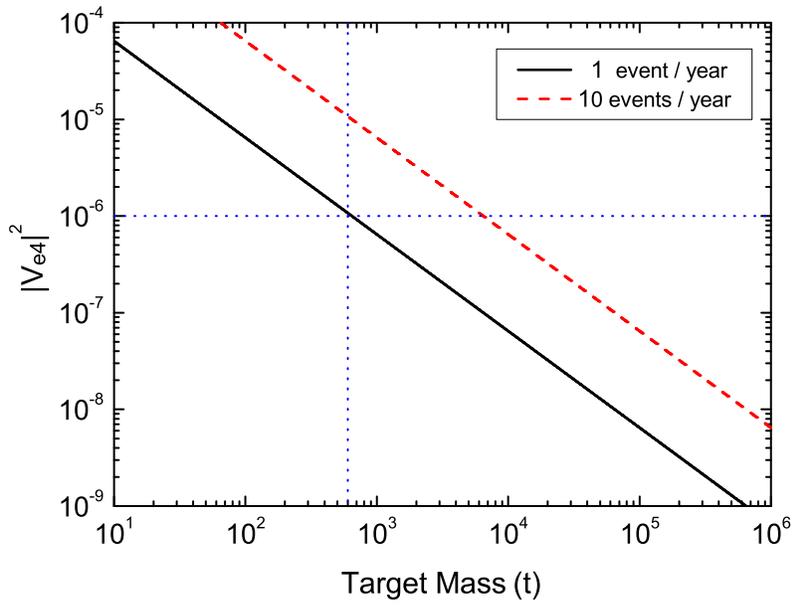}
\end{tabular}
\end{center}
\caption{The iso-rate curves of warm sterile antineutrino DM for the
mixing matrix element $|V^{}_{e4}|^2$ versus the target mass. Here
$n^{}_{\nu^{}_4} \simeq n^{}_{\overline{\nu}^{}_4} \simeq 5 \times
10^{4} \ (3 ~{\rm keV}/m^{}_4) ~{\rm cm}^{-3}$ in our Galactic
neighborhood has been input.}
\end{figure}
\end{document}